\documentclass[sn-mathphys,Numbered]{sn-jnl}

\usepackage{graphicx}%
\usepackage{float}
\usepackage{multirow}%
\usepackage{amsmath,amssymb,amsfonts}%
\usepackage{amsthm}%
\usepackage{mathrsfs}%
\usepackage[title]{appendix}%
\usepackage{xcolor}%
\usepackage{textcomp}%
\usepackage{manyfoot}%
\usepackage{booktabs}%
\usepackage{algorithm}%
\usepackage{algorithmicx}%
\usepackage{algpseudocode}%
\usepackage{listings}%
\usepackage[version=4]{mhchem}
\usepackage{lineno}
\usepackage[sectionbib]{bibunits}
\defaultbibliographystyle{unsrt} 
\defaultbibliography{references}

\begin{document}

\bibliographyunit[\chapter]




\title[Article Title]{Brillouin laser-driven terahertz oscillator up to 3~THz with femtosecond-level timing jitter}

\author*[1]{\fnm{Brendan} \sur{Heffernan}}\email{bheffern@imra.com}

\author[1]{\fnm{James} \sur{Greenberg}}\email{jgreenbe@imra.com}

\author[2]{\fnm{Takashi} \sur{Hori}}

\author[2]{\fnm{Tatsuya} \sur{Tanigawa}}

\author[1]{\fnm{Antoine} \sur{Rolland}}\email{arolland@imra.com}

\affil[1]{\orgdiv{Boulder Research Labs}, \orgname{IMRA, America Inc.}, \orgaddress{\street{1551 S. Sunset Street}, \city{Longmont}, \postcode{80501}, \state{CO}, \country{USA}}}

\affil[2]{\orgdiv{Japan Branch Office}, \orgname{IMRA, America Inc.}, \orgaddress{2-1 Asahi-machi}, \city{Aichi 448-8650}, \country{Japan}}


\abstract{The terahertz (THz) frequency range, spanning 0.1 to 10 THz, is a field ripe for innovation with vast, developing potential in areas like wireless communication and molecular spectroscopy. Our work introduces a dual-wavelength laser design that utilizes stimulated Brillouin scattering in an optical fiber cavity to effectively generate two highly coherent optical Stokes waves with differential phase noise inherently mitigated. To guarantee robust operation, the Stokes waves are optically injected into their respective pump lasers, which also serves to greatly improve the resulting coherence. The frequency difference between the two wavelengths is converted into THz waves through a uni-traveling-carrier photodiode. This innovative design facilitates the generation of THz waves with phase noise levels of less than -100 dBc/Hz, translating to timing noise below 10~$\mathrm{as} / \sqrt{\mathrm{Hz}}$ at 10 kHz Fourier frequency, over a carrier frequency range from 300 GHz to 3 THz. This development in phase noise reduction establishes a new benchmark in the spectral purity of tunable THz sources. Such advances are pivotal for applications to move beyond oscillator constraints.}


\maketitle


\section*{Introduction}

Terahertz radiation is currently being investigated for use in a wide range of applications\cite{Leitenstorfer2023}, including wireless communication \cite{Kurner2022}, radioastronomy~\cite{cliche2004high, nand2011ultra}, instrumentation for satellites~\cite{schoeberl2006overview}, electron accelerators~\cite{nanni2015terahertz}, and molecular spectroscopy~\cite{motiyenko2014rotational, swearer2018monitoring, wang2018chip, cuisset2021terahertz}. Applications such as wireless communications and molecular spectroscopy rely on continuous wave (CW) sources with a high degree of monochromaticity, and their performance metrics degrade with increasing source noise. For example, the stability of molecular clocks that use modulation spectroscopy is currently limited by phase noise \cite{Wang2021c}, highlighting the need for sources with lower phase noise.

Optical photomixing shows perhaps the greatest aptitude for providing low-noise and tunable THz waves \cite{alouini1998dual}. The low-noise aspect is evident in recent demonstrations that leveraged photomixing combined with optical frequency division (OFD) to demonstrate unprecedented low-phase noise generation of microwaves \cite{fortier2011generation,xie2017photonic} and 300 GHz waves \cite{tetsumoto2021optically}. However, the use of frequency combs for OFD limits the tunability to less than 1\% of the center frequency~\cite{fortier2016optically, zhang2019terahertz,wang2021towards,kuse2022low}. In the absence of OFD, photomixing two CW laser lines translates optical phase noise to a lower frequency, resulting in a THz wave with a spectral purity equal to the quadratic sum of the two lasers' interference on the photomixer. Using ultra-low linewidth lasers is a straightforward method of decreasing the THz phase noise \cite{Kittlaus2021}, but is ultimately limited by the uncorrelated nature of the two lines.

Correlating the phase noise of the two lasers can significantly decrease the linewidth/phase noise of the radiated THz. This can be achieved electronically via phase-lock-loops (PLLs), but PLLs have a very limited bandwidth (less than 100 kHz) and result in a servo bump. The phase noise of the resulting THz waves seldom matches that of leading multiplied microwave references. A promising alternative is to use a shared optical cavity resulting in two optical lines which are inherently correlated. This approach has been demonstrated directly \cite{Djevahirdjian2023} using dual-polarization solid-state lasers~\cite{rolland2014narrow} and monolithic semiconductor lasers~\cite{kim2009monolithic}. It has also been shown indirectly using optical frequency comb-disciplined laser diodes~\cite{quraishi2005generation, shin2023photonic}. However, this method has yet to be combined with very narrow optical linewidths.

Fiber lasers using stimulated Brillouin scattering (SBS) are renowned for achieving narrow linewidths~\cite{smith1991narrow}. SBS provides both optical amplification and filtering within a narrow bandwidth, e.g. 100 MHz in optical fiber at telecom wavelengths. Achieving SBS requires sustained, high-intensity light-matter interaction, commonly attained in microresonators~\cite{loh2015noise} or long fiber resonators with appropriate free spectral ranges~\cite{ducournau2011highly}. Importantly, SBS results in a Stokes laser line which is reduced in linewidth and phase noise compared to the pump via acoustic damping and optical feedback \cite{debut2000linewidth}. The degree of reduction depends primarily on cavity quality factor (Q). Typically, a linewidth narrowing by a factor of $\sim 10^3$ is achievable with a fiber cavity extending to hundreds of meters. From a pump laser with a 1~MHz linewidth, it is possible to generate a Stokes wave with a linewidth of 1~kHz. However, such a long fiber length results in a cavity's free spectral range (FSR) on the order of a few MHz, which is less than the gain bandwidth, thus leading to the potential for mode-hopping in the Stokes wave. The practical challenge, then, is to minimize the Stokes linewidth (thus increase the cavity length) while also finding a way to mitigate mode-hopping and retain manageable pump powers.

Past work has attained mode-hop free operation by limiting the cavity length to ensure the FSR is larger than the Brillouin gain bandwidth. The reduced cavity length necessitates pumping on resonance to achieve large intra-cavity powers, and subsequently requires locking the pump to the resonance through the Pound-Drever-Hall technique~\cite{loh2019ultra,li2019low}. Importantly, the amount of linewidth narrowing is limited due to the achievable cavity Q. Another solution is to increase the cavity length, pump non-resonantly, and phase-lock the pump-Stokes beatnote to a microwave reference \cite{danion2016mode}. In both cases, the feedback bandwidth is constrained by the capabilities of the pump laser actuators and, like the aforementioned PLLs, results in servo bumps which degrade the phase noise. The limited feedback bandwidth also necessitates a pump laser with high spectral purity and, consequently, a higher cost.

Here, we introduce a novel Brillouin fiber laser architecture that utilizes a pump laser injection locked to the generated Stokes line. This configuration eliminates locking electronics as well as its associated servo bumps, while also allowing the usage of pump lasers with only moderate linewidths~\cite{huang2016linewidth,huang2017tens,lopez2021application,alouini2021ultra}. Additionally, the architecture enables wide tunability. We use this laser in a dual-wavelength configuration as a source of low phase noise opto-THz waves. We adopt the term ``opto-THz wave'' to denote the optically carried THz wave prior to photomixing and radiation (the supplementary information provides a mathematical explanation of what constitutes an opto-THz wave). We present methods for measuring the phase noise of such a source in the THz and optical domains and find unmatched phase noise levels of less than -100 dBc/Hz for Fourier frequencies above 1~kHz for carrier frequencies between 300~GHz and 3~THz.

\section*{Dual-wavelength Brillouin laser as an opto-THz wave generator}

\begin{figure}[ht!]
    \centering\includegraphics[width=80mm]{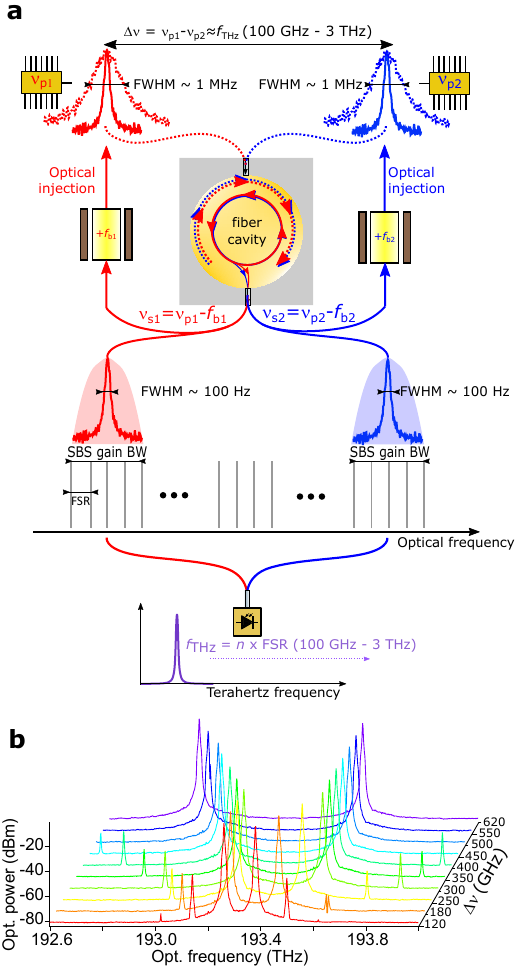}
    \caption{\label{figure1} \textbf{Schematic diagram of the DWBL and resulting spectra.} \textbf{a}: This illustration presents a conceptual perspective on the implementation of a DWBL as an opto-THz source that utilizes the stimulated Brillouin scattering effect within a non-resonantly pumped fiber cavity. FWHM: full width at half maximum; SBS: stimulated Brillouin scattering; FSR: free spectral range. \textbf{b:} A series of 10 measured optical spectra are depicted, showcasing the optical output and tunability of the local oscillator. These spectra represent the sweeping of the frequency difference between the two Stokes waves, ranging from 120~GHz to 620~GHz, while maintaining the identical diode laser pumps. Typical optical SNR ratios are 70~dB, measured in a 0.02~nm bandwidth.}
\end{figure}

The concept of the dual-wavelength Brillouin laser (DWBL) is illustrated in Fig.~\ref{figure1}a. In this setup, a pair of diode lasers operating at frequencies $\nu_{p1}$ and $\nu_{p2}$, with a spectral linewidth on the order of $1~\text{MHz}$, are utilized to pump a fiber cavity. The frequency difference between the two diode lasers is given by $\Delta\nu = \nu_{p1} - \nu_{p2}$ corresponding approximately to the target terahertz frequency output $f_{THz}$. This arrangement leads to the generation of two counter-propagating Stokes waves at optical frequencies $\nu_{s1}$ and $\nu_{s2}$, which are defined as $\nu_{s1,s2} = \nu_{p1,p2} - f_{b1,b2}$, where $f_{b1,b2}$ are the respective Brillouin frequency shifts.

In order to lock the phase and frequency of each pump through optical injection \cite{Liu2020a}, its respective Stokes wave frequency must be up-shifted by \( f_{b1,b2} \) to align with the pump wavelength. This task is accomplished using an electro-optic modulator (EOM). Although the modulator does not strictly ``shift'' the frequency, it induces optical sidebands at a distance from the Stokes signal corresponding to its modulation frequency. By modulating the EOM at \( f_{b1,b2} \), a sideband is produced at \( \nu_{s1,s2} + f_{b1,b2} = \nu_{p1,p2} \), and this sideband is subsequently injected back into the pump laser. This results in the elimination of mode-hopping, as the pump is continuously locked to the Stokes wave. 

Another important aspect of locking the pumps to the Stokes using optical injection is that it enables the system to approach the thermal noise floor of the fiber cavity. Provided that the spectral purity of the Stokes wave can be fully imparted onto the pump's spectral characteristics, an iterative narrowing of the Stokes linewidth is anticipated, progressing until the thermal noise boundary of the fiber cavity is reached. The capture range for optical injection in our configuration spans tens of megahertz, leading to cavity-determined phase noise out to large Fourier frequencies. Detailed descriptions of the fiber cavity configuration are provided in the methods section and the supplementary information (SI).

The resulting output is two, narrow-linewidth Stokes waves. These two optical tones have a high degree of phase noise correlation because they originate from the same fiber cavity. This makes the DWBL an attractive candidate for an opto-THz source, as the phase noise of the output waves cancel to first order upon photodetection. A uni-traveling-carrier photodiode (UTC-PD) can be used to convert the opto-THz wave into a terahertz wave \cite{Ishibashi1997}. Furthermore, the source exhibits extensive tunability, with $f_{br} = \nu_{s1} - \nu_{s2}$ spanning from 180~GHz to 620~GHz utilizing a single pair of pump lasers, as shown in Fig.~\ref{figure1}b. By switching to a different pair of diode lasers, the frequency difference can extend up to approximately 3~THz, primarily constrained by the bandwidth of certain optical components like erbium-doped fiber amplifiers in this study. The resilience of this architecture was demonstrated through continuous frequency counting of an opto-THz wave at 316~GHz, employing an electro-optic comb for down-conversion~\cite{rolland2011non}. The system routinely exhibits mode-hop free operation exceeding a few days (see supporting data in the SI).

\section*{Phase and timing noise metrology of terahertz waves}

\begin{figure}[ht!]
\centering\includegraphics[width=\textwidth]{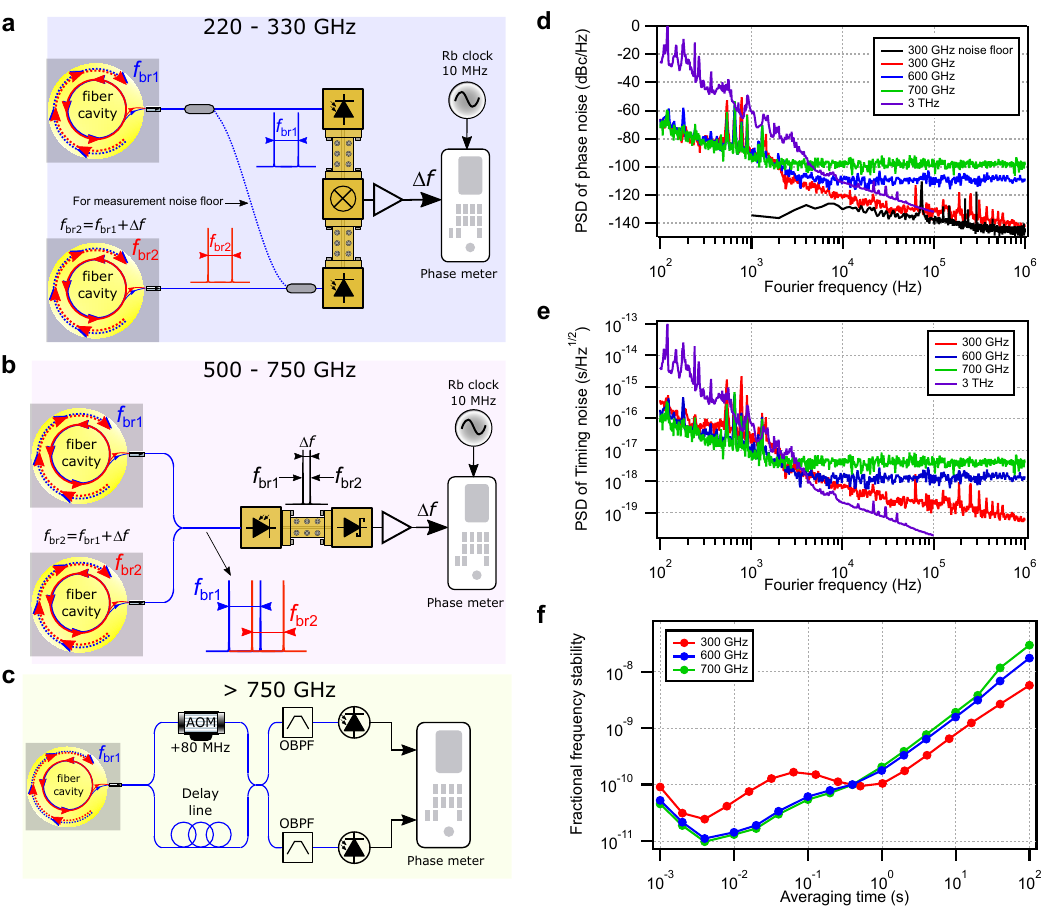}
\caption{\label{figure2} \textbf{Schematics for THz frequency metrology and results.} \textbf{a:}~Experimental implementation of a phase noise bridge to measure the spectral purity of a pair of oscillators at frequencies around 300 GHz using two UTC-PDs. \textbf{b} A second implementation for measuring frequencies between 500 - 750 GHz using one UTC-PD and one Schottky barrier diode (SBD). \textbf{c:}~ Illustration of the experimental configuration of a two-wavelength delay line interferometer to measure opto-THz phase noise. This setup is employed in situations where we did not have photodetectors to convert the optical signal into terahertz radiation, specifically when the frequency difference exceeds 750 GHz. AOM: acousto-optic modulator; OBPF: optical bandpass filters.\textbf{d:}~Power spectral density of phase noise of a pair of oscillators at 300~GHz~(red), 600~GHz~(blue), 700~GHz~(green), and 3 THz (purple) as well as the measurement noise floor at 300~GHz~(black). \textbf{e:}~Power spectral density of timing noise of a pair of oscillators at 300~GHz~(red), 600~GHz~(blue), 700~GHz~(green) and 3 THz (purple). \textbf{f:}~Fractional frequency stability in terms of Allan deviation of a pair of oscillators at 300~GHz~(red), 600~GHz~(blue), 700~GHz~(green). Error bars are smaller than the data point markers.}
\end{figure}

We conducted measurements on the power spectral density of the phase noise at four different frequencies: 300 GHz, 600 GHz, 700 GHz, and 3 THz (the SI shows amplitude noise at 300~GHz and 600~GHz). In initial measurements that compared the phase noise of the terahertz oscillator to a multiplied microwave reference (through an EO down-conversion)~\cite{rolland2011non}, it became apparent that the phase noise of the microwave reference restricted the measurement. To address this limitation, we duplicated the DWBL. This allowed for the measurement of phase noise from two independent but identical oscillators, which were not correlated with each other (see Methods for further discussion). By doing so, the constraint imposed by the microwave reference phase noise was overcome. Due to the wide range of opto-THz frequencies and availability of THz components, three different measurement schemes were used, shown in Fig.~\ref{figure2}a-c. 

In the 300~GHz-band, two UTC-PD were available as well as a fundamental mixer. WR3.4 waveguide-coupled components were used. We tuned the frequency difference of the Stokes lines in each oscillator to approximately 300~GHz. The two sources were then detuned by $\Delta f = $~100~MHz. The light from each DWBL was incident on its own independent UTC-PD (see Fig.~\ref{figure2}a). The UTC-PDs were driven at 9~mA photocurrent emitting approximately 150 µW of radiated power (see SI for power measurements) and were waveguide-coupled to a mixer. The fundamental mixer output was a radiofrequency signal at $\Delta f$, which served as an intermediate frequency that carried the combined phase noise of the two terahertz oscillators. This intermediate frequency phase noise was compared with an RF oscillator with an absolute phase noise lower than the 300~GHz phase noise. 

In the 600~GHz frequency band, only one UTC-PD was available. Therefore, we changed the phase noise measurement setup to that shown in Fig.~\ref{figure2}b. Two opto-THz oscillators were tuned to approximately 600~GHz. The two Stokes lines from each oscillator were combined in the optical domain through a fiber coupler (optical spectra shown in the SI). Two pairs of Stokes waves (at 600~GHz and 600~GHz+$\Delta f$) were sent to the UTC-PD. The UTC-PD converted this optical signal into two terahertz waves at 600~GHz and 600~GHz+$\Delta f$. The radiated power was estimated to be 10 µW. A single Schottky barrier diode coupled to a WR1.5 waveguide detected the beatnote envelope between the two terahertz tones, acting as a low-pass filter and generating an RF signal at $\Delta f$. Similar to the setup at 300~GHz, the signal $\Delta f$ contained the phase noise of both terahertz sources and can be measured against an RF reference. 

For opto-THz frequencies larger than 750 GHz, a UTC-PD was not available in our lab and therefore phase noise could not be measured directly in the THz domain. As an alternative, we evaluated the phase noise using the experimental setup depicted in Fig.~\ref{figure2}c. The setup is based on a two-wavelength delayed interferometer (TWDI) \cite{Kuse2018}. The opto-THz wave is essentially measured against an optical fiber delay line. 

Phase noise results are shown in Fig.~\ref{figure2}d (300~GHz (red), 600~GHz (blue), 700~GHz (green) and 3~THz (purple)). We present the phase noise of two oscillators, i.e., 3~dB was not subtracted from the data (see Methods). From 300~GHz to 700~GHz, we measure a phase noise level of -70~dBc/Hz at 100 Hz while we measure it at -20~dBc/Hz at 3~THz. This apparent increase in phase noise at a 3 THz operation point is caused by limitations on the measurement itself (see supporting data in the SI). Therefore, the phase and timing noise for 3 THz presented in Fig.~\ref{figure2}d-e should be considered an upper estimate, i.e. the actual noise is lower than what was measured.  White phase noise levels, as expected, increase with higher carrier frequencies due to less emitted terahertz power. At 2~kHz Fourier frequency, the phase noise level is -98~dBc/Hz from 300~GHz to 700~GHz. Above this Fourier frequency, the 700~GHz phase noise is dominated by the -98~dBc/Hz white phase noise. The 600~GHz signal reaches its white phase noise of -110~dBc/Hz at 4~kHz Fourier frequency. At 300~GHz, a much lower white phase noise floor was observed, since we had more emitted power and had access to a fundamental mixer with only 10~dB of conversion loss. A phase noise level of -120~dBc/Hz is observed at 10~kHz Fourier frequency and reaching down to -145~dBc/Hz at 1~MHz Fourier frequency. Using a shared opto-THz oscillator to drive the two UTC-PDs in the 300GHz configuration depicted in Fig.~\ref{figure2}a, we were able to simultaneously assess the residual phase noise of both UTC-PDs and the fundamental mixer (referred to as the ``measurement noise floor'' in the legend of Fig.~\ref{figure2}d). Importantly, we confirmed that the presence of $1/f$ flicker phase noise in the UTC-PDs and the fundamental mixer did not impede the evaluation of the Brillouin laser spectral purity. 

The calculated timing noise is presented in Fig.~\ref{figure2}e. This quantity is proportional to phase noise, but is normalized by the carrier frequency, making it a good metric of comparison across many frequencies. Because technical noise should cancel to first order in the photodetection process, we expect that timing noise will decrease with a higher carrier frequency. Fig.~\ref{figure2}e confirms this prediction. Specifically, when observing Fourier frequencies ranging from 100Hz to 1kHz, it becomes evident that the spectral purity at 700GHz is fractionally superior compared to that at 300GHz. This trend does not hold for 3 THz, again, due to limitations on the phase noise measurement itself (see SI). Regardless, the measured timing noise level is below 1~fs/$\sqrt{Hz}$ for Fourier frequency exceeding 300~Hz at all carrier frequencies from 300~GHz to 3~THz. The integrated timing noise (timing jitter) across the measured Fourier frequency range are as follows: 2.317 fs at 300 GHz, 1.412 fs at 600 GHz, 3.359 fs at 700 GHz, and 26.15 fs at 3 THz.

\begin{figure}
    \centering\includegraphics[width=\textwidth]{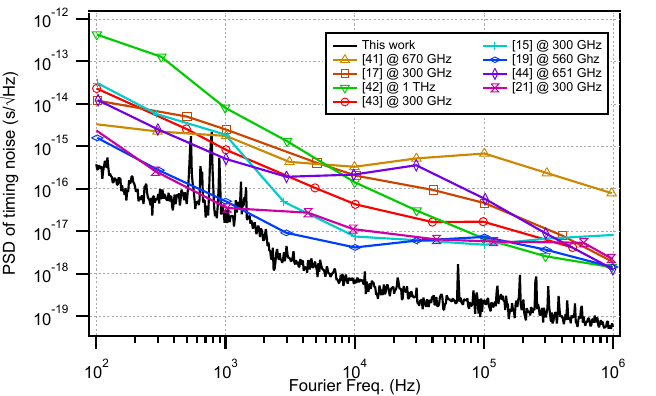}
    \caption{\label{tn_comp} \textbf{Comparison of THz sources' timing noise.} A plot of timing noise for state-of-the-art THz sources operating at frequencies ranging from 300 - 1000 GHz.}
\end{figure}

For carrier frequencies of 300-700 GHz, phase noise analysis was performed within a Fourier frequency range of 100 Hz to 1 MHz. Although the fractional phase noise at these carriers was small, the absolute fluctuations were restrictively noisy when observed from the RF domain. Consequently, measuring phase noise for Fourier frequencies below 100 Hz became a challenge because of the tracking bandwidth of the phase noise analyzer. To measure fluctuations below 100~Hz, we opted to measure the absolute frequency difference ($\Delta f$) against a rubidium standard (as shown in Fig.~\ref{figure2}a-b) and then calculated the Allan deviation to determine the fractional frequency stability for averaging times ranging from 1~ms to 100~s. The results are presented in Fig.~\ref{figure2}f. We used a zero dead-time frequency counter with a 1~ms gate time. This allowed us to overlap with the phase noise measurement and to look in the 10~ms to 1~s range. The thermal floor of the Brillouin cavity was reached at 4~ms averaging time. Peak stability was $2.4 \times 10^{-11}$, $1.12 \times 10^{-11}$, and $1.05 \times 10^{-11}$, for carrier frequencies at 300~GHz, 600~GHz and 700~GHz, respectively. Similar to observations of the timing noise measurements, fractional frequency stability improved for higher carrier frequencies up to approximately 400ms, where random walk processes became dominant. Beyond 400ms, the fractional frequency stability was better for lower carrier frequencies. At an averaging time of 100~s, the carrier frequency experienced a drift of a few kHz.

\section*{Conclusions and outlook}

In conclusion, our research contributes significantly to the expanding field of terahertz (THz) technology, particularly in the development of low-noise and tunable sources. We have designed and presented a dual-wavelength Brillouin laser, where the pump lasers are injection-locked to the Stokes waves. When used as an opto-THz source, the single fiber cavity ensures highly-correlated phase fluctuations between the two optical lines, while the injection locking provides mode-hop-free operation and iterative linewidth reduction. Following photodetection, the oscillator demonstrates exceptional phase noise performance at 300~GHz across the entire measured Fourier frequency range. We compare the timing noise at 300 GHz with other state-of-the-art sources between 300 - 1000 GHz in Fig.~\ref{tn_comp} \cite{DeSalvo2012a, Lampin2020, zhang2019terahertz, tetsumoto2021optically, tetsumoto2020300, Wang2022, kuse2022low, Djevahirdjian2023}. Note that most of the oscillators in this comparison have no clear path for implementing wide tunability. Our result constitutes an unprecedented timing noise level in the terahertz domain. The achieved phase noise levels, recorded at below -100 dBc/Hz at a 10 kHz Fourier frequency, set a new standard in the field. This is particularly noteworthy given the extensive bandwidth of the system, which extends up to 3 THz.

While our oscillator showcases record-breaking phase noise performance, there are still aspects ripe for imminent improvement. By employing optical frequency division with this novel laser, as outlined in a microcomb approach \cite{tetsumoto2021optically}, there is potential to achieve extraordinary phase noise levels at millimeter-wave frequencies. Theoretically, reaching phase noise as low as -140~dBc/Hz at a 10 kHz Fourier frequency could become feasible, although this depends upon further advances in phase noise metrology within this frequency range. Beyond 1 THz, where the power emitted by UTC-PDs falls short and high-power oscillators typically exhibit broadened linewidth, this source could effectively seed a quantum cascade laser to enhance the terahertz power output with superior spectral purity~\cite{freeman2017injection}. Another area for further enhancement is the management of frequency drift, a common characteristic of most oscillators. Maintaining frequency stability over extended periods is crucial, particularly for practical applications in real-world scenarios. One promising approach to enhance this stability is to synchronize the oscillator with an atomic clock, known for its exceptional long-term stability. Additionally, investigating native frequency references to the THz domain could provide an innovative method to further solidify the frequency stability of these sources. If this source were to be synchronized with a rotational molecular reference, an intermodulation effect~\cite{audoin1991} might define an upper limit for its stability, potentially achieving less than $1 \times 10^{-14}/\sqrt{\tau}$ at 3~THz. This level of stability would be comparable to that of compact atomic clocks, presenting an exciting prospect in terms of performance.

These results, to our understanding, set a new benchmark for tunable and low noise sources in the THz domain, enabling a new realm of possibilities for applications such as wireless communication and molecular spectroscopy. We expect this work to act as a catalyst, motivating further research and innovations in this rapidly progressing field.


\section*{Acknowledgments}
We express our gratitude to Tadao Nagatsuma of Osaka University for providing us with the 600 GHz UTC-PD and detector. The second dual-wavelength Brillouin laser was built using funding provided through the National Institute of Information and Communications Technology (NICT), Japan, Grant No. 00901.


\section*{Methods}

\subsection*{Brillouin laser cavity}
The two optical lines are generated from a 75m-long fiber cavity. Stimulated Brillouin scattering is used as a gain medium. Two isolator-free C-band diode lasers (Eblana Photonics, EP1550 Discrete-Mode Series) spectrally separated by the target terahertz frequency are used to initiate Brillouin gain. They are amplified to a total power of approximately 200~mW using an erbium-doped fiber amplifier (EDFA). The gain is centered at a frequency down-shifted by $\sim$11 GHz from each respective pump frequency. The two pump diode lasers are not resonant with the fiber cavity, which consists of an optical circulator, 75 m of polarization-maintaining fiber, and a 95/05 fiber output coupler (see SI for diagram). However, the two Stokes waves are resonant with the cavity. The cavity quality factor was experimentally measured to be $8 \times 10^{8}$. 

To ensure mode-hop-free laser operation, we use a new technique of optically injecting each Stokes wave to its respective pump laser diodes. To match the pump frequencies, the two Stokes waves frequencies are independently up-shifted by the Brillouin frequency shift via an electro-optic modulator driven by a synthesizer. Determining the Brillouin frequency shift is realized via the photodetection of the optical beatnote between the pump and its associated Stokes wave. This mechanism guarantees monomode oscillation of the Stokes waves. Additionally, because the stimulated Brillouin scattering coupled to acoustic damping and cavity feedback reduces the linewidth of the pump laser diodes, iteration of this effect leads to a frequency noise equilibrium between the pump and the Stokes. This means that the Stokes wave frequency noise reaches the thermal noise of a 75-m fiber cavity. Since optical injection is separately realized on each pump laser, tuning one laser frequency can be done independently of the other by simply adjusting the synthesizers used to shift the Stokes waves. Therefore, it is possible to tune the laser diodes' frequencies to reach a defined terahertz frequency. A piezoelectric device is integrated within the optical fiber cavity to enable fine-tuning of the opto-THz frequency.

The cavity is enclosed inside a vacuum chamber at 100 mTorr. This allows a temperature stability of the fiber spool to be on the order of +/-1mK. To isolate the cavity from mechanical vibrations, it is housed within a vacuum chamber, which is then placed on a benchtop vibration isolation platform (specifically, a Minus K negative stiffness vibration isolator, model 50BM-10). This entire setup is mounted on a rack.  

\subsection*{Phase noise measurements at 300, 600, and 700 GHz}
An Agilent Technologies PXA signal analyzer (N9030A) was used to measure phase noise. Prior to measurement, the RF beatnote (in the range of $\sim$ 100 MHz) was divided by 32 to acquire a signal with smaller absolute phase noise for more precise quantification. The resulting phase noise was then scaled to the correct value by adding $20\log(32)$ (in units of dBc/Hz). 

Generally, in order to measure the phase noise of an oscillator, a lower noise oscillator is required. If there is none, an identical oscillator must be built and the two compared. When two identical oscillators are used, their individual phase noises contribute quadratically to the observed total phase noise in the measurement system. This is because phase noise, being a random process, combines in a root-sum-square manner when two independent noise sources are involved. In the context of a pair of oscillators, the measured phase noise is theoretically expected to be 3 dB higher than the phase noise of a single oscillator due to this quadratic addition. This assumes the two oscillators are truly identical. In principle then, a subtraction of 3 dB could be applied to estimate the phase noise of an individual oscillator.

The phase noise data presented in the main text \textit{has not} had this factor of 3 dB removed, and is therefore a more conservative estimate of the phase noise of a single Brillouin opto-THz oscillator.

\newpage

\bibliographyunit[\chapter]

\section*{Supplemental}
\subsection*{Opto-THz waves}
Here we provide a derivation of THz phase noise when produced through photomixing as well as provide mathematical formalism for the concept of opto-THz waves. In classical electromagnetism, light is characterized by its electric field vector $E$. The orientation of this vector determines the direction of light polarization~\cite{born1999principle}. Because electromagnetic waves are transverse, the electric fields $E_1$ and $E_2$ of monochromatic plane waves can be described as:

\begin{equation}
    E_{1,2}\cos(\omega_{1,2}t-k_{1,2}z).
\end{equation}

\noindent
Let $k$ denote the wave vector, which, for simplicity, is aligned along the z-axis. We drop this spatial dependence to focus more clearly on the frequency and phase components. The electric field of the total opto-THz wave can be expressed as:
\begin{equation}
\label{Equ1}
    E(t)=E_1 \cos(\omega_1 t-\phi_1)+E_2 \cos(\omega_2 t-\phi_2),
\end{equation}

\noindent
where $E_1,E_2$ are the constant amplitudes of the two optical lines. The phase of each field is considered to be a constant, time-averaged value with time-dependent fluctuations about the mean such that $\phi_{1,2} = \Phi_{1,2} + \varphi_{1,2}(t)$, where $\Phi_1$ and $\Phi_2$ are constants. Derived from equ.~\ref{Equ1}, the Poynting vector $\Pi$ of such optical electric field is written as follows:
\begin{equation}
\label{Equ2}
    \Pi(t)\propto \mid E(t) \mid ^2 = E_1^2\cos^2(\omega_1t-\phi_1)+E_2^2\cos^2(\omega_2t-\phi_2)+2E_1E_2\cos(\omega_1t-\phi_1)\cos(\omega_2t-\phi_2)
\end{equation}

\noindent
Equation~\ref{Equ2} can be rewritten as:

\begin{equation}
\label{Equ3}
\begin{aligned}
    \Pi(t) &\propto \mid E(t) \mid ^2 \\
    &= \frac{E_1^2}{2}[1+\cos(2\omega_1t-2\phi_1)]\\
    &\quad +\frac{E_2^2}{2}[1+\cos(2\omega_2t-2\phi_2)]\\
    &\quad +E_1E_2[\cos((\omega_1-\omega_2)t-(\phi_1-\phi_2))\\
    &\quad\qquad +\cos((\omega_1+\omega_2)t-(\phi_1+\phi_2))].
\end{aligned}
\end{equation}

\noindent
The Poynting vector, as described in equ.~\ref{Equ3} has a slow component with an angular frequency of $\omega_1-\omega_2$ and a phase of $\phi_1-\phi_2$. It is this component that is responsible for optically carrying the THz wave.  To convert an opto-THz wave to THz radiation, the two electric fields that make up the opto-THz wave are mixed on a photodiode.  Uni-traveling-carrier photodiodes (UTC-PD's) provide effective conversion. Using UTC-PD's, radiation up to 3~THz have been observed~\cite{ishibashi2021uni}. The average optical power $P_{opt}$ over the photodiode time constant $\tau_{PD}$ ($\mid\omega_1-\omega_2\mid\ll1/\tau_{PD}\ll\omega_1,\omega_2$) is:

\begin{equation}
    \label{equ4}
    P_{opt}(t)\propto E_1^2+E_2^2+2E_1E_2\cos((\omega_1-\omega_2)t-(\phi_1-\phi_2))
\end{equation}

\noindent
The photocurrent $I_s$ through the photodiode load $R_{load}$ can be expressed as:

\begin{equation}
\label{equ5}
    I_s(t)=\eta E(t)
\end{equation}

\noindent
with $\eta$, the photodiode responsivity. The emitted terahertz power $P_{THz}$ through this photodetection process can be written as follows:
\begin{equation}
    P_{THz}=\langle R_{load} \times I_s^2(t)\rangle \propto R_{load} \times P_{opt}(t)
\end{equation} 

\noindent
The phase of the opto-THz wave is $\phi_1-\phi_2$. As such, the phase fluctuations of the opto-THz $\varphi_{THz}$ will be dominated by the phase fluctuations $\varphi_1$ and $\varphi_2$ of the two optical lines as: 
\begin{equation}
    \varphi_{THz}(t)=\varphi_1(t)-\varphi_2(t).
\end{equation}
\noindent
When phase fluctuations $\varphi_1(t)$ and $\varphi_2(t)$ are strongly correlated, it can significantly reduce the opto-THz wave phase fluctuations, as these correlated fluctuations cancel upon photomixing. However, photomixing introduces its own phase noise~\cite{quinlan2023photodetection}. This includes flicker noise $\varphi_{1/f}$, referred to as 1/f noise, as well as thermal and shot white phase noise $\varphi_{TandS}$. Finally, we can write the terahertz signal as such:
\begin{equation}
    E_{THz}(t)=V_{THz}[1+\alpha_{THz}(t)]\cos[\omega_{THz}*t+\varphi_{THz}(t)+\varphi_{1/f}+\varphi_{TandS}]
\end{equation}
\noindent where $V_{THz}$ is the time-averaged field amplitude and $\alpha_{THz}(t)$ is the time-dependent fluctuation of the amplitude, which is assumed to have a time-averaged value of 0.

\subsection*{Dual-wavelength Brillouin laser architecture}

\begin{figure}[ht!]
\centering\includegraphics[width=\textwidth]{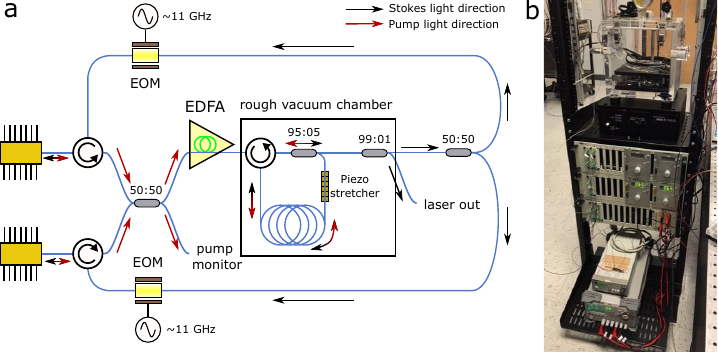}
\caption{\label{brillouin} \textbf{Dual-wavelength Brillouin architecture.} A schematic overview of the laser is shown in a. Black arrows indicate the direction of the Stokes light while red arrows indicate the direction of the pump light. A photograph of a rack-mounted iteration of the laser is shown in b.}
\end{figure}

\subsection*{Continuous, mode-hop-free operation}

\begin{figure}[ht!]
\centering\includegraphics[width=\textwidth]{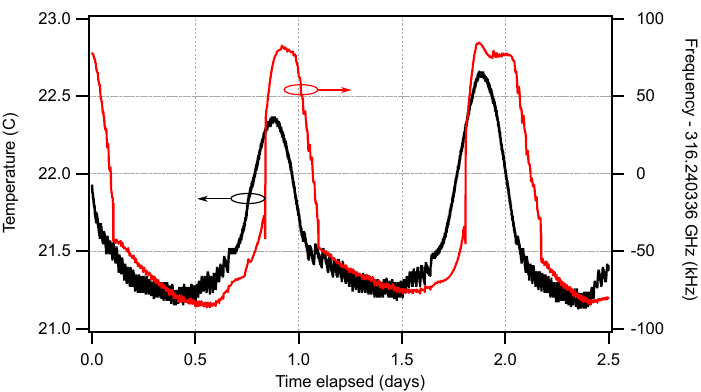}
\caption{\label{modehop} \textbf{Dual-wavelength Brillouin opto-THz output with time.} The opto-THz frequency output of the dual-wavelength Brillouin is plotted in red. The frequency changes with room temperature, but does not mode hop over the 2.5 days of measurement time.}
\end{figure}

We continuously monitored the output of the dual-wavelength Brillouin source for over 48 hours using EO-comb frequency readout \cite{rolland2011non}. The results are plotted in Fig. \ref{modehop}. The output frequency drifts with room temperature, also plotted in Fig. \ref{modehop} (black). Clear day-night cycles can be observed. While there are relatively fast frequency drifts, on the order of 5 minutes, frequency changes are less than 100 kHz from the average, far less than the $\sim$ 2.4 MHz FSR of the cavity indicating mode-hop-free operation over more than 2 days.

\subsection*{THz power and amplitude noise}

\begin{figure}[ht!]
\centering\includegraphics[width=\textwidth]{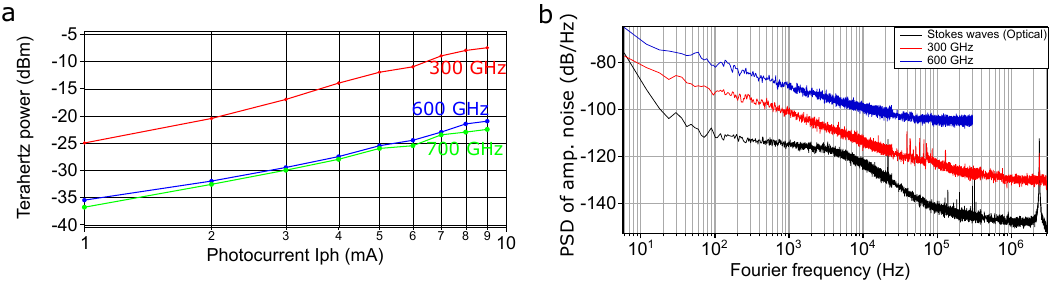}
\caption{\label{rin} \textbf{UTC-PD power output and Brillouin RIN.} \textbf{a:} Radiated THz power as a function of UTC-PD photocurrent. \textbf{b:} Power spectral density of the relative intensity/amplitude noise of the Brillouin opto-THz source measured after photomixing at 300 GHz (red), after photomixing at 600 GHz (blue), and in the optical domain (black).}
\end{figure}

In Fig.\ref{rin}a, we present a graph delineating the relationship between the radiated power and the measured photocurrent from a uni-traveling-carrier photodiode (UTC-PD). Since the UTC-PD is coupled to a millimeter-wave waveguide, it is noteworthy that the UTC-PD deployed for the 300 GHz measurements differs from the one utilized for the 600 and 700 GHz assessments. Likewise, two Schottky barrier diodes were used; one for the measurements at 300 GHz, and one for the measurements at 600 and 700 GHz. The specified responsivity for the 300~GHz photodiode is 0.24~A/W. For the UTC-PD at 600 GHz, the responsitivity is about 0.14~A/W.

The UTC-PDs operating at 300 GHz and 600 GHz are interfaced with WR-3.4 and WR-1.5 rectangular metallic waveguides, respectively. These configurations are meticulously designed to ensure optimal functionality within their corresponding frequency bands of 300 GHz and 600 GHz. A broadband coupling circuit, monolithically integrated with the photodiode chip, facilitates the transition between the UTC-PD and the waveguide. The constructed module exhibits a 3-dB bandwidth of hundreds of GHz, indicative of its robust operational capacity across a substantial portion of the terahertz spectrum. Specifically, the photomixer's frequency span at 300 GHz covers 200 GHz to 400 GHz, while at 600 GHz, it extends from 450 GHz to 900 GHz.

To empirically demonstrate the impact of emitting lower power, we conducted measurements of the power spectral density of amplitude noise at two distinct terahertz frequencies: 300 GHz and 600 GHz, as depicted in Fig. \ref{rin}b. The methodology involved a straightforward coupling of a single Schottky barrier diode, serving as an amplitude detector, with the UTC-PD. For the 300 GHz carrier, a white noise floor at -130 dB/Hz was observed, whereas the 600 GHz carrier displayed a noise floor at -110 dB/Hz. Additionally, flicker noise, characterized by a 1/f frequency dependence, was detected, which is postulated to originate from the Schottky barrier diode; however, this aspect was not explored within the scope of this study. Similarly, the potential saturation effects of the Schottky diode at 300 GHz, which could present a deceptively lower Relative Intensity Noise (RIN) in comparison to the 600 GHz scenario, were not investigated. Across all frequencies, the measured RIN of the radiated terahertz waves exceeded that of both Brillouin lines quantified in the optical domain

\subsection*{Brillouin optical spectra}
\begin{figure}[ht!]
\centering\includegraphics[width=\textwidth]{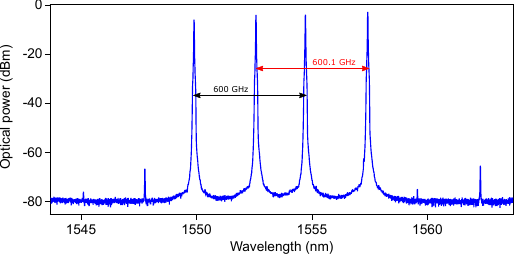}
\caption{\label{brillouin_600_spectra} \textbf{Combined optical spectra of two Brillouin lasers.} This configuration is used to measure the phase noise in the WR1.5 band. The optical signal to noise on each line is greater than 70 dB.}
\end{figure}

The optical spectrum of two separate dual-wavelength Brillouin systems, combined through a 50:50 coupler, is shown in Fig. \ref{brillouin_600_spectra}. This configuration was used for phase noise measurements at 600 and 700 GHz, as detailed in the main text. When this spectrum undergoes photodetection, two terahertz waves, at 600 GHz and 600.1 GHz are emitted. These waves are then down-converted to baseband with the use of a Schottky diode.

\subsection*{Measurement of 3 THz phase noise and limitations of TWDI}



\begin{figure}[ht!]
\centering\includegraphics[width=\textwidth]{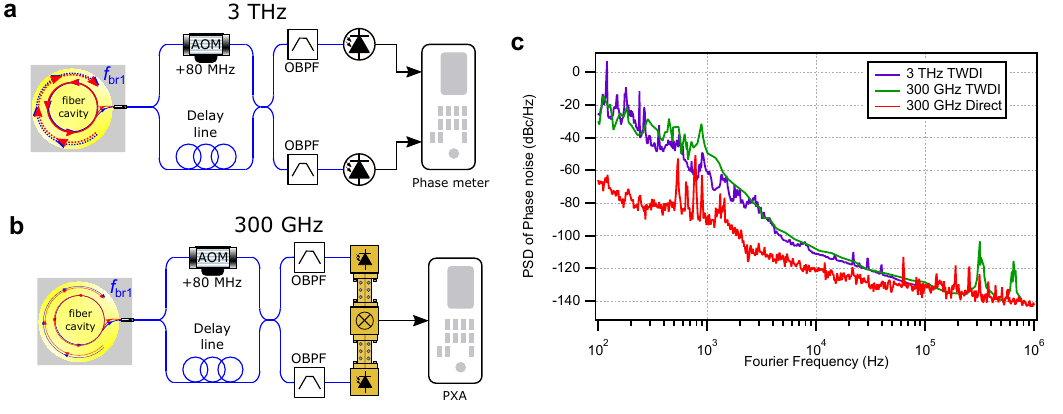}
\caption{\label{twdi} \textbf{Phase noise measurements at 3 THz and their limitations.} Phase Noise Characterization at THz Frequencies. (a) illustrates the TWDI setup for 3 THz phase noise measurement, and (b) for 300 GHz, with the latter using UTC-PD's for wave radiation. (c) shows that similar phase noise results (purple for 3 THz and green for 300 GHz) suggest limitation by fiber noise in the delay line. The more accurate 300 GHz phase noise (red), as detailed in the main text, indicates that TWDI provides only an upper limit for phase noise, especially at 3 THz.}
\end{figure}

We conducted phase noise measurements at 3~THz using a two-wavelength delayed interferometer (TWDI) as shown in Fig. \ref{twdi}a. The result of this measurement is plotted in Fig. \ref{twdi}c. 

In order to assess the limitations of this measurement, we also characterized the phase noise of 300 GHz radiation using our TWDI. However, for 300 GHz measurements, the opto-THz wave was converted to THz using two separate UCT-PDs and subsequently mixed on a fundemental mixer, as shown in Fig. \ref{twdi}b.

The results of these experiments are summarized in Fig. \ref{twdi}c. Notably, the TWDI measurements for both 3 THz (illustrated in purple) and 300 GHz (shown in green) exhibit nearly identical phase noise profiles. This parallelism strongly implies that the phase noise is predominantly influenced by the fiber noise introduced by the delay line, as referenced in~\cite{Kuse2018}, and this phenomenon constrains the measurement accuracy at both tested frequencies—0.3 and 3 THz.

A more sensitive measurement of phase noise at 300 GHz, detailed within the main body of the text and duplicated here in Fig.\ref{twdi}c (red line), corroborates that the limitations of the TWDI method affected our ability to accurately quantify phase noise at 300 GHz. When these observations are considered collectively, they lead to the conclusion that the TWDI methodology, as implemented in our experiments, is not sufficiently sensitive to ascertain the true phase noise at 3 THz. Instead, our findings represent an upper boundary of the phase noise, underscoring the need for enhanced measurement techniques to fully characterize the noise profile at this high frequency.

\bibliography{bib}



\end{document}